\documentstyle[sprocl,psfig]{article}

\def\be{\begin{equation}}
\def\ee{\end{equation}}
\def\bea{\begin{eqnarray}}
\def\eea{\end{eqnarray}}

\begin{document}

\title{AMBIGUITIES IN DETERMINATION OF SELF-AFFINITY  IN THE
 AE-INDEX TIME SERIES}

\author{N. W. WATKINS, M. P. FREEMAN, C. S. RHODES\footnote{Currently at: 
  DAMTP, Centre for Mathematical Sciences, Wilberforce Road, Cambridge, CB3 0WA}}

\address{ British Antarctic Survey, High Cross, 
Madingley Road,\\ Cambridge, CB3 0ET, UK \\E-mail: NWW@bas.ac.uk} 

\author{ G. ROWLANDS}

\address{ Space and Astrophysics Group, University of Warwick\\
Coventry, CV4 7AL, UK 
\\E-mail: phscv@snow.csv.warwick.ac.uk }


\maketitle\abstracts{  \begin{center} {\bf Abstract} \end{center}
The interaction between the  Earth's magnetic field and the solar wind plasma
results in a natural plasma confinement system which stores energy. 
Dissipation of this energy
 through Joule heating in the ionosphere 
can be studied via the Auroral Electrojet (AE) index.
The apparent broken power law form of the frequency spectrum of 
this index  has motivated investigation of whether it can be described as
fractal coloured noise.  One frequently-applied test for self-affinity is to demonstrate
linear scaling of the logarithm of the structure function of a time series with
the logarithm of the dilation factor $\lambda$.  
We point out that, while this is conclusive when applied to signals that are  
self-affine over many
decades in $\lambda$, such as  Brownian motion, the 
slope deviates from exact linearity and the conclusions become
ambiguous when the test is used over shorter  ranges of $\lambda$. We demonstrate that  
non self-affine time series made up of random pulses can show near-linear
scaling over a finite dynamic range such that they could be misinterpreted
as being self-affine. In particular we show that pulses
  with functional forms such as those identified by Weimer  
 within the $AL$ index, from which $AE$ is partly derived, will exhibit nearly linear scaling over ranges
  similar to those previously shown for $AE$ and $AL$. The value of the slope, related to the Hurst
  exponent for a self-affine fractal, seems to be a more robust discriminator 
  for fractality, if other information is
  available. }

\section{INTRODUCTION}


The characterisation of global energy transport in the coupled solar
wind-magnetosphere-ionosphere system is a fundamental problem in  space
plasma physics~\cite{ke}. Solar wind energy is transferred to, stored by,
and ultimately released from the magnetosphere by a range of mechanisms, in
which substorms play a key role. Most investigations of the substorm problem
have focused on single substorms or small groups of similar events,
analogous to the study of individual earthquakes in seismology.

A complementary approach is to analyse inputs to and outputs from the system
in an attempt to constrain the range of possible physics occurring in the
magnetospheric ``black box'' (c.f. analogous approaches in climatology and
seismology~\cite{tu}). Reviews of the significant progress made so far in
applying the methods of low dimensional chaos to the magnetosphere are given
by Klimas {\it et al}~\cite{kl} and Sharma~\cite{sh}; while more recent
investigations into whether or not the ``black box'' can be treated as a
self-organised critical (SOC) system~\cite{je} are reviewed by Watkins {\it %
et al}~\cite{wat00}, Chapman and Watkins~\cite{chap00} and Consolini and
Chang~\cite{con00}. One mechanism for dissipation of magnetospheric energy
is through Joule heating in the ionosphere's auroral electrojets. This
process can be studied via the auroral electrojet ($AE$) index, which is a
means of estimating the electrojet current. The Joule energy dissipated
depends upon both this and the ionospheric conductivity. $AE$ is available
at 1-minute resolution. Tsurutani {\it et al}.~\cite{ts} showed this to have
a ``broken power law'' frequency spectrum. The high frequencies
approximately follow $f^{-2}$ while the lower frequencies are $f^{-1}$ with
a break at about $1/5$ h$^{-1}$. Power law frequency spectra are common in
nature and can have several causes~\cite{je} such as Kolmogorov turbulence
or the bifurcation route to chaos. They are thus in themselves not
sufficient to completely constrain simple models. A parallel effort to
studies of the power spectrum has been the search for low dimensionality,
initially through the Grassberger-Procaccia (GP) algorithm~\cite{kl,sh}.
However, as noted by Osborne and Provenzale~\cite{op}, a low and fractional
GP dimension is not uniquely a signature of low dimensional chaos. It is
also compatible with self-affine coloured noise~\cite{op} or SOC~\cite{ch}.
In view of the fact that $AE$ is known {\it a priori} to be the output of a
complex system, Takalo and Timonen, in an important series of papers$^{12-16}
$, 
investigated whether the dynamics of magnetospheric and auroral indices were
better encapsulated by stochastic ``coloured noise'' rather than by chaos.
One test applied to $AE$~\cite{tt94} was for self affinity - a property of
both coloured noise and chaos. A particularly important technique for
identifying self-affinity in the work of Takalo and Timonen$^{12-16}$ was
the use of the second order structure function $S_{2}$ (although other
methods have also been applied to this problem$^{17-19}$). In this paper, by
constructing a simple example, we illustrate that $S_{2}$ alone cannot
reliably distinguish exponential autocorrelation from intrinsic
self-affinity in the short timescale part of the $AE$ signal, which has been
linked to the substorm ``unloading'' timescale~\cite{tt98}. By considering
how $S_{2}$ is related to other measures of self-affinity we address the
question of what additional knowledge may be required to make $S_{2}$ more
useful.

\begin{figure}[t]
\psfig{figure=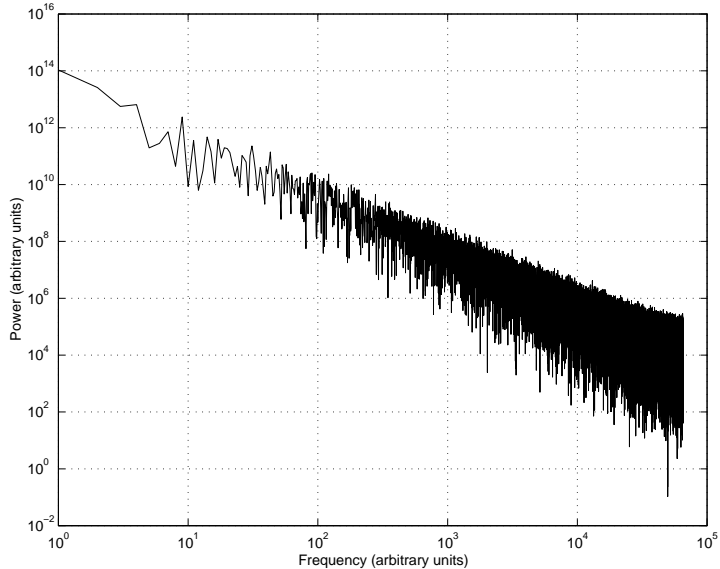,height=3.0in}
\caption{ Power spectrum of model Brownian motion ($H=0.5$). }
\label{fig:radish}
\end{figure}

\section{SELF-AFFINITY ($H=0.5$) IN BROWNIAN MOTION}

There are two kinds of fractal: self-similar and self-affine~\cite{ro}. They
are distinguished  by  whether the rescaling necessary to produce  the
original object is isotropic (self-similar) or anisotropic (self-affine). 
In the case of a random fractal such as a time series $X(t)$, one is testing
for statistical rather than exact self-affinity, so the test applied~\cite%
{tt94}  uses the second order structure function~\cite{tt94,ro} $S_2
(\lambda)$, defined by

\begin{equation}
S_2 (\lambda) = < ( X(t + \lambda \Delta t) - X (t) )^{2} >
\end{equation}
where $< ...>$ denotes an average over time $t$. For a self affine curve $%
X(t)$, 
\begin{equation}
S_2(\lambda) \sim \lambda^{2H} S_2 (1)
\end{equation}
where $H$ is the Hurst exponent  ($0 < H < 1$ for a self-affine fractal) and 
$S_2(1) = < (X(t+ \Delta t) - X(t))^{2} >$~\cite{ro}.  This results in
linear dependence  (with slope $H$) of $\log \ [ S(\lambda) / S(1) ]^{1/2}$
on the logarithm of the dilation factor $\lambda$. We note that not only is
it not necessary for $\lambda$ to be small~\cite{tt94}  but that
self-affinity in fact implies that the above holds for all scales $\lambda$.
The time stationarity assumption implicit in equation (1)~\cite{ro} allows
us to use the definition of the normalised autocorrelation function $ACF
(\lambda \Delta t)$:

\begin{equation}
ACF (\lambda \Delta t) = \frac{< (X(t + \lambda \Delta t) X(t) > }{<
X^{2}(t) >}
\end{equation}

to rewrite $S_2(\lambda)/S_2(1)$ in terms of the ACF

\begin{equation}
\frac{S_2(\lambda)}{S_2(1)} = \frac{(1 - ACF(\lambda \Delta t))}{(1 -
ACF(\Delta t))}.
\end{equation}

\begin{figure}[t]
\psfig{figure=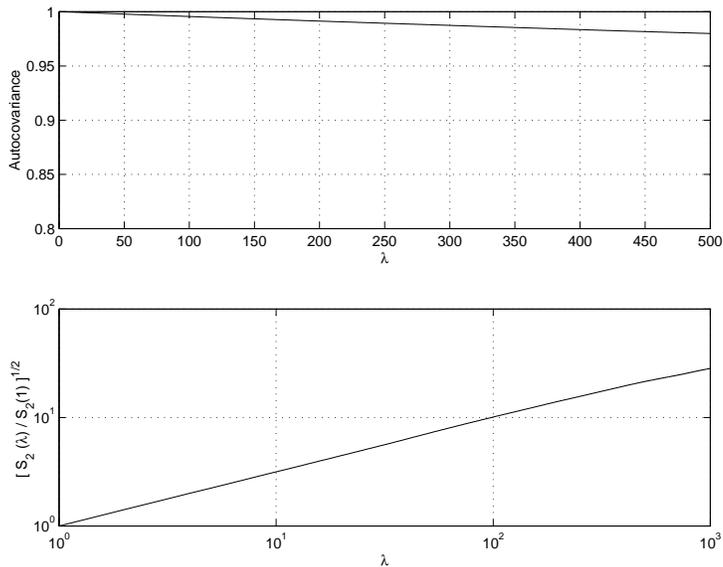,height=3.0in}
\caption{Autocovariance (top panel) and scaling plot (bottom panel) for
model Brownian motion ($H=0.5$). }
\label{fig:radish}
\end{figure}
Alternatively one may form the numerator and denominator of (3) from the
time-averaged, time-lagged, products of the series $\delta X = X(t) - \bar{X}
$ (see equation (1) of Takalo and Timonen~\cite{tt94}). We follow
engineering convention~\cite{be86} in referring to equation (3) with $\delta
X$ replacing $X$ as  the normalised autocovariance (ACV). Equation (4) holds
with $ACV(\lambda \Delta t)$ replacing $ACF(\lambda \Delta t)$, so either
can be used as a test for self-affinity~\cite{tt94}. In numerical work we
will follow Takalo and Timonen~\cite{tt94a} in using the ACV. It is
calculated for a discrete series ($X_i$; $i=1, ..., N$ with mean $\bar{X}$)
by 
\begin{equation}
ACV(j) = \frac{\sum_{i=1}^{N-j} ( X_{i}-\bar{X} ) (X_{i+j}-\bar{X})}{%
\sum_{i=1}^{N} (X_{i}-\bar{X})^2}.
\end{equation}

A classic example of a process which is both self-affine and fractal is
Brownian motion. Figure 1 shows a representative power spectral estimate
(unwindowed periodogram) for a time series of  131072 points of simple
Brownian motion ($H=0.5$).  The well-known $f^{-2}$ form is easily seen,
limited only by the available dynamic range of the data.  The upper panel of
figure 2 shows the normalised autocovariance of  the same time series.

The lower panel of figure 2 shows $\log [S_2(\lambda) / S_2(1)]^{1/2}$
versus $\log \lambda$, where we calculate $S_2$ using the normalised
autocovariance from equation (5). The range in the plot of $S_2$ was chosen
for ease of comparison with figure 4 of Takalo and Timonen~\cite{tt94} and
our figure 6. The curves in both panels of figure 2 are nearly straight
lines. The value $H=1/2$ can be read off from the slope of the line in the
lower panel of figure 2. As expected, the structure function is an effective
detector of its original intended target, a wide spectrum self-affine
fractal signal.

\section{APPARENT SELF-AFFINE FRACTALITY ($H=0.5$) IN EXPONENTIALLY
CORRELATED RANDOM PULSES}

The identification problem of self-affinity over a finite range begins to be
apparent when one applies the structure function method to a series of
random pulses. We first consider the case of random time series which have
exponential autocorrelation function.  Many physically interesting random
processes can be well approximated by an exponential ACF~\cite{be}. As an
exactly soluble example we note the simple ``random telegraph"~\cite{je,be}.
This is a two level Poisson-switched process which switches between level $F$
and level $-F$ with constant probability $1/\tau$ per unit time. This
process has~\cite{je,be} an autocorrelation function:

\begin{equation}
ACF (\lambda \Delta t) \sim e^{-2 | \lambda \Delta t| / \tau}
\end{equation}
which, by the Wiener-Khinchine theorem, indicates  a power spectrum of the
form $f^{-2}$ for high frequencies  ($f \gg 1/\tau$), but flat ($f^{0}$) for
low frequencies ($f \ll 1/\tau$)~\cite{je}.  Because $e^{-2 | \lambda \Delta
t| / \tau}  = 1 - 2 | \lambda \Delta t| / \tau + {\mbox O} (\lambda^2
\Delta^2 t^2)$  the scaling of $\log \sqrt{S_2(\lambda) / S_2(1)}$ versus $%
\lambda$ will not only be  linear (i.e. apparently self-affine) for $\lambda
\Delta t$ small compared with $\tau/2$ but will also give rise to a Hurst
exponent value of $1/2$ if $H$ is derived from the slope of the line (i.e.
apparent fractality).

Without knowing {\em a priori} that it is a 2-level, Poisson-switched 
system, application of $S_2$ to a time series that was exponentially
autocorrelated over time could cause one to  infer (erroneously) that the
short lag behaviour corresponding to times $\lambda \Delta t < \tau/2 $ was
both self-affine and fractal. This serves to underline the point that
self-affinity is an intrinsically wide bandwidth property, and that
application of a wide-band test over the restricted range ($\lambda \Delta t
< \tau/2$) makes it hard to distinguish certain types of randomness from
self-affine fractality.

\section{APPARENT SELF-AFFINITY IN WEIMER PULSES.}

The relevance of the above observations to the $AE$ time series becomes
clearer when we consider that $AE$ contains recurring ``pulses" associated
with magnetospheric substorms. Both the pulse shape and its recurrent
properties could give rise to the observed scaling in $AE$. We first
consider apparent scaling due to the pulse itself, and then examine the
behaviour of a random series of such pulses.

\subsection{Restricted range self-affinity from a single Weimer pulse}

The pulse shape was studied by Weimer~\cite{we} in the AL index,  one of the
two indices from which $AE$ is derived ($AE=AU-AL$).  A random sample of 55
substorms was divided into three classes based upon  the peak $AL$ value
attained. For each class, the $AL$ time series were  superposed with respect
to the substorm epoch, from which the average time  series was then
calculated. The three resultant average substorm profiles  were shown to be
well fitted by the functional form $\alpha p t e^{-pt}$  with both $\alpha$
and $p$ increasing with increasing peak $AL$. This  functional form is the
solution of an ordinary second-order differential  equation that was argued
to describe the evolution of the electric field  and currents in the
substorm current wedge. The ionospheric part of the  substorm current wedge
is a westward current that the $AL$ index was  designed to measure.

We now show that this shape causes apparent scaling in $S_2$ at small values
of $\lambda \Delta t$ in the case of a single, isolated Weimer pulse. We
take $\alpha =1$ without loss of generality. The numerator ($ACF^{*}$) of
equation (3) becomes: 
\begin{equation}
ACF^{*} (\lambda \Delta t) = < pt e^{-pt} p (t + \lambda \Delta t) e^{-p (t
+ \lambda \Delta t)} >.
\end{equation}

By starting with the identity 
\begin{equation}
< e^{-2 p t} > = \int_0^{\infty} e^{-2pt} dt,
\end{equation}
we may evaluate averages such as (7) by differentiation with respect to $p$.
We find 
\begin{equation}
ACF^{*} (\lambda \Delta t) = \frac{1}{4p} (1 + p \lambda \Delta t) e^{-p
\lambda \Delta t}
\end{equation}
and so using the denominator of (3) to normalise the ACF we have 
\begin{equation}
ACF (\lambda \Delta t) = (1 + p \lambda \Delta t) e^{-p \lambda \Delta t}
\end{equation}
Expanding the normalised ACF as a Taylor series gives 
\begin{equation}
ACF (\lambda \Delta t) = 1 - \frac{1}{2} p^2 \lambda^2 \Delta t^2 +
O(\lambda^3)
\end{equation}
which yields, on insertion into the right hand side of equation (4), a
scaling of $S_2(\lambda) / S_2(1)$ with $\lambda^2$, for $\lambda \Delta t$
small compared with $1/p$ (observed to be $\approx 30$ minutes). This
implies linear behaviour when the logarithm of either $S_2(\lambda) / S_2(1)$
or its square root is plotted against $\log \lambda$. Hence the pulse then
appears self affine over this range with $H=1$.

\subsection{Restricted range self-affinity from random Weimer pulse train}

Now let us investigate the scaling properties of a sequence of such pulses,
as might occur in the $AE$ time series when measured, for example, over the
100 days (144000 points) studied by Takalo and Timonen~\cite{tt94}.  As in
the random telegraph we chose a random sequence of pulses, specialised here
to  a representative example  of the Weimer pulse shape. Each pulse was of
form $\alpha p t e^{- p t} $ where $p= 1/30$ minutes$^{-1}$, $\alpha=1$, and
the sampling interval was $1$ minute for 131136 points. The inter-pulse
intervals were drawn from an exponential distribution with e-folding time $%
300$ minutes$^{24}$.  The above model is not meant to provide an exhaustive
model for the $AE$ time series, but the pulse is a known~\cite{we} component
of the $AL$ (and thus $AE$) signal and so its contribution to the apparent
self-affinity of $AE$ must be investigated.

\begin{figure}[t]
\psfig{figure=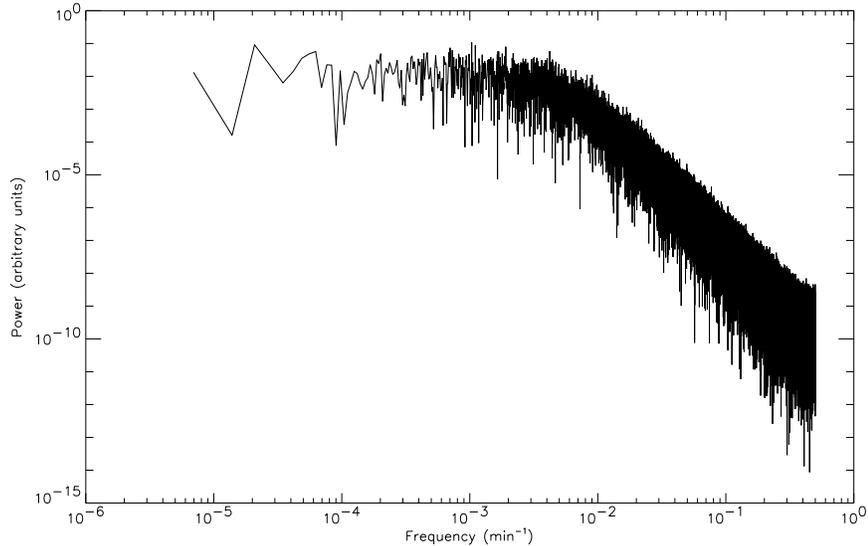,height=3.0in,angle=90}
\caption{Representative example of a spectrum from a random Weimer pulse
train. }
\label{fig:radish}
\end{figure}

\begin{figure}[t]
\psfig{figure=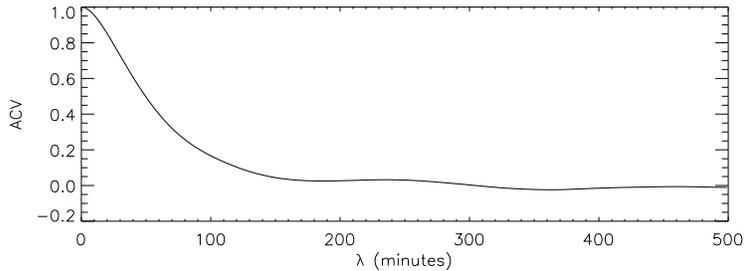} \vskip 0.5cm
\caption{ Autocovariance of random Weimer pulse train }
\label{fig:radish}
\end{figure}

\begin{figure}[t]
\psfig{figure=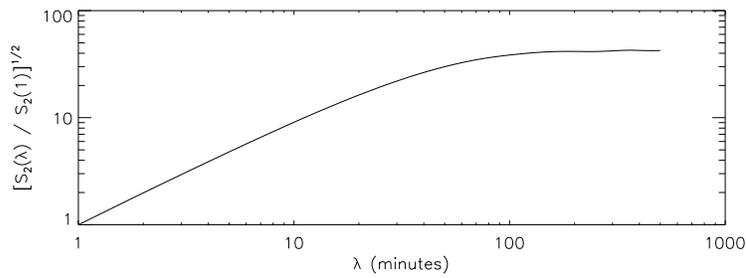} \vskip 0.5cm
\caption{Scaling plot for random Weimer pulse train. }
\label{fig:radish}
\end{figure}

\begin{figure}[t]
\psfig{figure=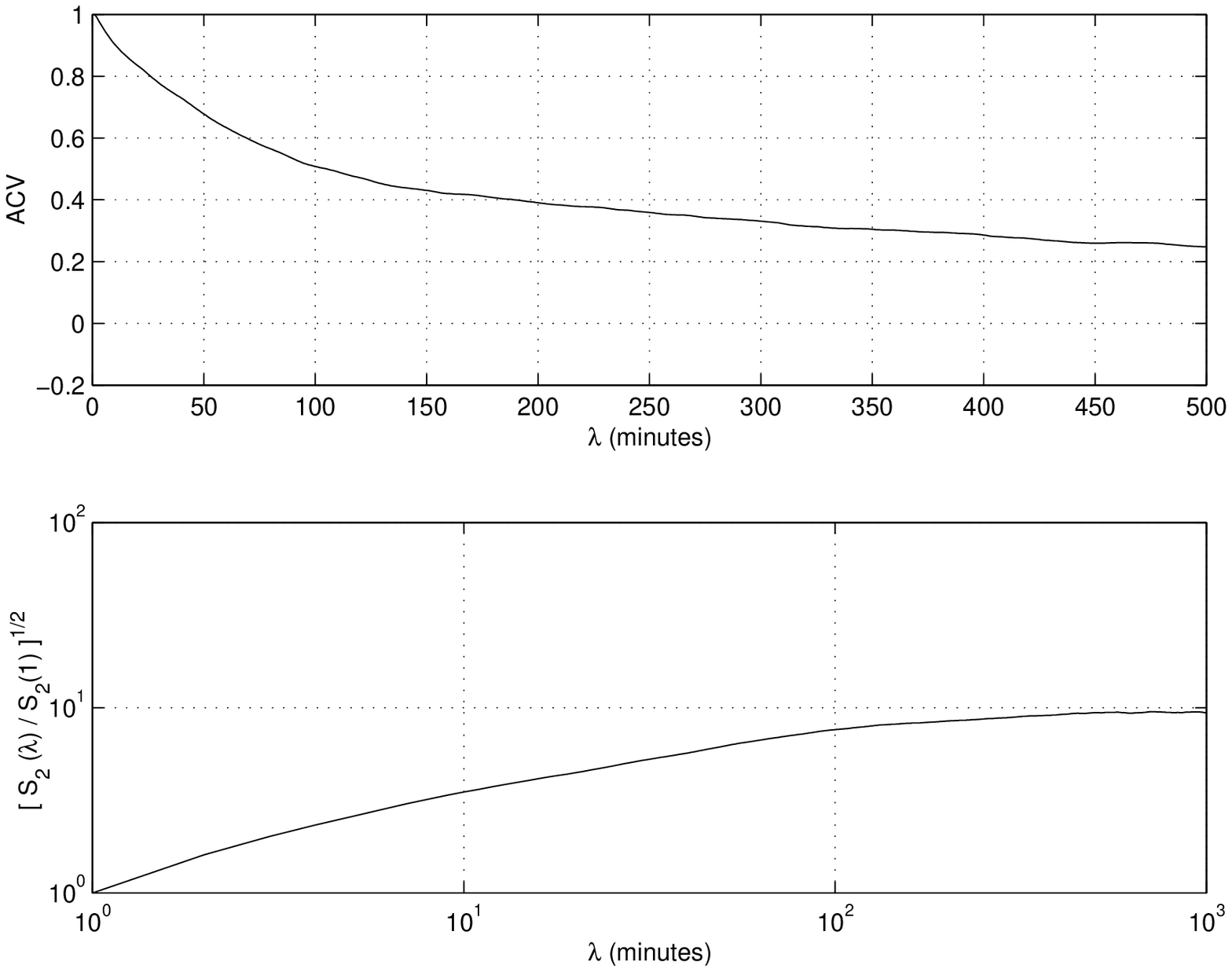,height=3.0in}
\caption{Autocovariance and scaling plot for 100 days of AE, starting on 1st
January 1983 }
\label{fig:radish}
\end{figure}

Figure 3 shows a spectrum estimate for the model time series. The spectrum
has the characteristics of the exponentially autocorrelated random telegraph
with a breakpoint at around $1/p$ between $f^0$ for $f \ll 1/p$ and $f^{-4}$
for $f \gg 1/p$. The time series gives rise to an autocovariance function
with a steep (quadratic) slope at small lags $\lambda \Delta t < 30 \ {\rm %
min} (= 1/p)$ (see figure 4) characteristic of the pulse shape. The
associated structure function has slope $\approx 1$ for $\lambda \Delta t$
less than 10 minutes, and progressively less than 1 as $\lambda \Delta t$
increases, such that it appears nearly linear over two decades in $\lambda
\Delta t$ (figure 5).

Again this near-linearity, used alone without other  information on a
natural signal of necessarily restricted dynamic range, could lead one to
infer self-affine properties (or indeed chaotic ones) in a signal that is
not self-affine. The addition of randomness to the single-pulse behaviour
described in section 4.1 has given rise to a Hurst exponent less than 1,
when measured over the whole of the range $1 < \lambda \Delta t < 100$. We
believe there to be competition between the effects of randomness (e.g. $%
H=0.5$ in the random telegraph) and the integer value of $H=1$ associated
with individual  pulses.

\section{$AE$ RE-EXAMINED, DISCUSSION AND CONCLUSIONS}

We now consider the scaling properties of the measured $AE$ time series in
the light of the previous examples. The top panel of figure 6 shows the
autocovariance of the first 100 days of $AE$ for 1983, and may be compared
with the top panel of figure 4 of Takalo and Timonen~\cite{tt94}. Again, the
steep (exponential) fall of the ACV results in a near linear slope for small 
$\lambda \Delta t$, and a slow decrease in the slope as larger and larger
ranges of $\lambda \Delta t$ is considered . Importantly, however, the slope
is, always less than 1 (J. Takalo, Private communication, 1999). Overall it
resembles near-linear scaling in the structure function for most of the
first two decades of $\lambda$ (bottom panel of figure 6), and (plotted in
the the middle panel of figure 4 of Takalo and Timonen~\cite{tt94}), was
cited by Takalo and Timonen~\cite{tt94} as a key piece of evidence for
self-affinity in $AE$. They also noted the resemblance of the $AE$
autocovariance function to an exponential and proposed that the
autocorrelation time of $AE$ be defined as the lag for timescales longer
than that over  which the autocovariance ceased to be exponential.
Inspection of figures 4, 5 and 6 lead us to conclude, however, that, unlike
the ideal case of Brownian motion, {\bf neither} the curve of $S_2$ for $AE$%
, {\bf nor} that of the simplified random Weimer pulse train are straight
over the range $\lambda \Delta t=1$ to $\lambda=120$. We remark that,
insofar as the ACF of $AE$ is exponential for small $\lambda \Delta t$,
there {\em must} eventually be a departure from near-linearity in the
structure function as $\lambda \Delta t$ increases, unless the range over
which the exponential behaviour is seen is so small that a straight line
would be just as good an approximation as the exponential.

In addition, both $AE$ and the model Weimer pulse train of section 4.2 give
a fractional $H$ value when taken over the whole range from $10$ to $100$.
Without {\em a priori} additional knowledge, we might equally well have
concluded that the random pulse train was self-affine over the range $%
\lambda \Delta t < 100$, but by construction we know this is not so.

Our model was deliberately simplified. In the natural $AE$ time series, the
extended tail of the ACV is  expected to reflect the solar wind-driven
component (also present in $AU$ and $AL$), which our simulation neglected.
As originally conjectured by Tsurutani {\it et al}~\cite{ts} the solar wind
driver is probably the origin of the ``1/$f$" part of the $AU/AL/AE$
spectrum~\cite{fre00,tak00}.

We may summarise our findings as follows. By construction of an explicit
counter-example we have shown that near-linear scaling of $S_2$ over about
two decades is not in itself sufficient to show self-affinity. We have also
given analytic and numerical evidence that non-fractal random series can
produce non-integer Hurst exponents over limited dynamic ranges. We thus
infer that self-affinity in the range $0$ to $100$ minutes for $AE$ has not
been and could not be proved by the use of $S_2$ alone.

One may reasonably point out that several other methods have been used to
provide evidence of self-affinity in geomagnetic indices; both in the papers
of Takalo et al. and those of other workers$^{17-19}$. One
may thus enquire as to what kind of additional knowledge or analysis
techniques would be necessary for considering the results of the structure
function method fruitful? Based on what we have found, we suggest that
answer is at least threefold.

1) {\em Be aware that many tests for fractality are actually designed assuming a fractal signal}: A test based on the assumption of fractality can disprove fractality but
cannot prove it. The methods for measuring fractal dimension that we are
aware of assume self-affinity in their design  i.e. they typically examine
the scaling behaviour of a signal. Only if they find no evidence of scaling
at  all is there no ambiguity.

2) {\em Use more than one test}: Several tests are better than one because different methods are sensitive
to different  non-fractal effects. Thus use of several tests means that a
series  with non-fractal aspects is less likely to be misinterpreted.  Most
of the methods for measuring fractal dimension which have been applied to
geomagnetic data are of one of two basic types. The first type of method
basically estimates the dimension of a fractal curve by examining how the
average value of short lengths of curve 
\begin{equation}
S_1 = < X(t + \lambda \Delta t) - X(t)>
\end{equation}
scales with the ruler length $\lambda$ (in units of the sampling interval $%
\Delta t$). Such methods have been applied by V\"{o}r\"{o}s~\cite{vor90} to
magnetometer data, and more recently to geomagnetic and solar wind
quantities by Price and Newman~\cite{price01}, who used the related,
cumulative "R/S" analysis.

The second type studies the positive definite second order function 
\begin{equation}
S_2 = <(X(t + \lambda \Delta t) - X(t))^2>
\end{equation}
and returns the same information~\cite{ro} as the ACF when estimated on a
stationary signal (see section 2).  For this reason it is thus also formally
related to the power spectrum via  the Wiener-Khinchine theorem. $S_2$, the
ACF and the power spectrum have all been extensively investigated for the $AE
$, $DSt$ and related indices by Takalo {\em et al}$^{12-16}$ .  The meaning of this family of techniques can
be understood as studying the behaviour of the histogram of variance of the
signal (or the power spectrum) with increasing time dilation $\lambda \Delta
t$ (or frequency); depending on whether one is dilating in time (in the case
of $S_2$ and the ACF) or frequency (in the case of the power spectrum). We
caution that time lag in the ACF or in $S_2$ is not trivially 1/(the Fourier
frequency) because any frequency in a Fourier transform has contributions
from multiple ACF lags and vice versa (see Bendat and Piersol~\cite{be86},
pages 120-122). In the case of a simple fractal, the dimension (and Hurst
exponent $H$) estimated from such methods should theoretically be the same
as from $S_1$, although in practice the errors of the two methods need not
be the same~\cite{san92}. If they differ substantially, this  may be a
pointer that the time series is not intrinsically a wideband fractal, and
that one of $S_1$ or $S_2$ is more sensitive to this.

An example of how additional tests for fractality have supplied new
knowledge is in the continuing  study of the $AE$ index. This has been known
since the work of Tsurutani {\em et al}~\cite{ts} to have a  ``1/$f$" low
frequency and ``1/$f^2$" high frequency power spectral density. Acting only
on information from the power spectrum or other $S_2$-type methods, one
might thus infer that AE is a bi-affine quantity$^{12-16}$, i.e. it has two separate scaling regions and a
break between them. In contrast, Consolini~\cite{con97} has
studied the ``burst distributions"~\cite{con99,con97} of $AE$. These are the histograms of
intervals between threshold crossing times (burst and inter-burst lifetimes)
and of areas above threshold between crossings (burst sizes). The lifetime
distributions are an $S_1$-type measurement~\cite{addison,fre00b} and were found to
have (exponentially rolled-off) scaling with a single slope over a very wide
range, interrupted only by a non-scaling component at about 2 hours. The
apparently paradoxical observation of bi-affine behaviour in $S_2$ and
``contaminated" mono-affine behaviour in $S_1$ has been addressed in two
different ways. One has been to introduce models which have the required
properties in both $S_1$ and $S_2$, such as forest fire models~\cite{con99}
or coupled map lattices~\cite{tak00}  driven by wideband solar-wind like
signals. The other, informed by the fact (section 4 and 5 above) that the
high frequency $f^{-2}$ part of a power spectrum {\em need not arise from a
fractal aspect of the time series}, has been to postulate~\cite{fre00} that
the $AE$ series is in fact a hybrid time series with a fractal element
arising from the solar-wind driven ionospheric current systems and a
non-fractal part arising from energy storage and release in the
magnetosphere (substorms). This was supported~\cite{fre00} by the
observation that the scaling in $AE$ (and $AU/AL$) burst lifetimes is the
same as that seen in the solar wind (see also Freeman et al~\cite{fre00b})
while the non-scaling component was seen only in the magnetospheric
quantities such as  $AU$ and $AL$~\cite{fre00}.

There have been exceptions to the use of $S_2$ or $S_1$ type techniques in
the geomagnetic context. We are grateful to an anonymous referee for
reminding us of the results of a multifractal analysis of the $AE$ index by
Consolini et al.~\cite{con96}. These results must imply some constraints on
possible models describing the variability of auroral currents. However, in
the same way that $AE$ when measured over 1 year by a method of type $S_1$
is essentially fractal~\cite{con97}, and required the use of several years'
measurements for the ``bump"-like feature in the otherwise scaling $S_1$ to
become apparent~\cite{fre00}, it seems to us that one might expect a
multifractal  analysis of less than 2 months of $AE$~\cite{con96} to give a
good fit to a p-model of turbulence because the solar wind driver is also
well fitted by this particular turbulence model~\cite{horbury}. We believe
that a  study on a much longer series of $AE$ would be required to exclude
even our own toy model of random differentiable Weimer pulse trains, when
superposed on the multifractal solar wind background. 
We note that, independently, a recent multifractal
study of geomagnetic data from Thule, Alaska has excluded the biaffine coloured noise model~\cite{voros00} for that dataset.

3) {\em Remember that Nature does not have to be purely fractal any more than it has to be non-fractal}: Many types of natural signal have both fractal {\bf and} non-fractal
components. In consequence, when using methods to examine fractality, one should be aware that it is possible to find something between the extremes of wideband fractality and none at all, as  discussed in point (2) above  for the case of $AE$. Another example is to imagine looking out of one's tea-room window at a tree through a regularly spaced window
blind. The distinguishing of the fractal tree and the periodic blind is a
task that the human eye and brain perform routinely, and which a Fourier
transform can also do because it can resolve the blind spacing as a spatial
frequency. A ``random blind" appearing at Poisson-switched intervals would
be much more of a problem for an FFT, and would be analogous to the pulses
of section 3 and 4. The user thus needs to determine how much the presence
of a ``contaminating" signal or signals in the fractal time series may
affect their interpretation, at which point the question may become as much
physical as mathematical. This is currently an admittedly very difficult
task because of the sparsity  of literature on such hybrid time series, and
is one which we plan to examine in more detail in future papers.

\section*{Acknowledgments}

We thank Joe Borovsky, Tom Chang, Sandra Chapman, Richard Dendy and Dave Willis 
for useful discussions, and the World Data Centre C1 at RAL for supplying 
the $AE$  index. We are grateful to Jouni Takalo for comments based on a careful
reading of the first version of the manuscript, and to an anonymous referee
for drawing our attention to references 17-19 and for
 several other useful
 suggestions. This  paper is based on a talk presented at the EGS/AGU NVAGA 4 conference,
Roscoff, 1998.

\end{document}